\documentclass[11pt]{article}
\usepackage[american]{babel}
\usepackage{amsmath}
\usepackage[hyperfootnotes=false]{hyperref}
\usepackage{verbatim}
\usepackage{amsfonts}
\usepackage[multiple]{footmisc}
\usepackage{cite}

\usepackage{fullpage}

\begin{document}

\begin{center}

\centerline{}
\centerline{}
\centerline{}
\centerline{}
{\bf \huge{\centerline{Constraints on the R-charges of Free Bound}
\centerline{
States from the R\"omelsberger Index}}}
\medskip

\renewcommand{\thefootnote}{\fnsymbol{footnote}}
\vspace{1.1 cm} {Efrat Gerchkovitz}\footnote{efrat.gerchkovitz@weizmann.ac.il}\\
\renewcommand{\thefootnote}{\arabic{footnote}}
\addtocounter{footnote}{-1}

\vspace{0.8 cm}{\it Department of Particle Physics and Astrophysics,\\
Weizmann Institute of Science, Rehovot
76100, Israel}\\

\vspace{0.8cm}

\vskip2pt
\end{center}

\noindent
The R\"omelsberger index on $S^3\times{\mathbb{R}}$ serves as a powerful test for conjectured dualities, relying on the claim that this object is an RG-invariant. In this work we support this claim by showing that the singularities suggested by Witten of ``states moving in from infinity'' are excluded on $S^3\times{\mathbb{R}}$. In addition, we provide an application of the R\"omelsberger index, in the form of a constraint on the RG flow of supersymmetric theories. The constraint, which applies for asymptotically free theories with unbroken supersymmetry and non-anomalous R-symmetry, is the following: if the R-charges of the chiral multiplets in the UV theory are $q_i\in(0,2)$ and the IR theory can be described as a free theory of chiral bound states, then the R-charges of these bound states, $\tilde{q}_j$, are constrained such that $\tilde{q}_j\in(0,2)$. We thus provide a proof of a weak version of a conjecture proposed by Intriligator. We mention some applications of this result.

\newpage

\section{Introduction}

Supersymmetric field theories on curved manifolds have become an active field of study in recent years, since new observables turned out to be exactly calculable on curved spacetimes \cite{Romelsberger:2005eg,Kinney:2005ej,Pestun:2007rz,Dolan:2008qi,Closset:2012vg,Closset:2013vra,Festuccia:2011ws,Dumitrescu:2012ha,Dumitrescu:2012at}.
One such observable is a Witten-like index on $S^3\times{\mathbb{R}}$, which is sometimes referred to as the ``superconformal index.'' This index was first introduced in \cite{Romelsberger:2005eg} and independently in \cite{Kinney:2005ej}. In the latter it was defined for superconformal theories, where the index can be interpreted as a partition function on $\mathbb{R}^4$ {in radial quantization}.
Since we do not limit ourselves to superconformal theories, we will refer to the index as ``the R\"omelsberger index.''

Like the Witten index, the R\"omelsberger index is independent of continuous deformations of the theory. Therefore, if the theory is continuous in the coupling constants, the index is independent of the couplings and, in particular, an RG-invariant. Independence of the couplings is the property that makes the index useful: it allows one to calculate the index in the free theory and by that to capture the behavior of a non-perturbative theory without having to worry about the complications caused by the interactions.

The R\"omelsberger index is used as a test for conjectured dualities: since the index was introduced, it was computed for many pairs of dual theories, and the results always matched \cite{Romelsberger:2007ec,Dolan:2008qi,Spiridonov:2008zr,Spiridonov:2009za,Spiridonov:2011hf,Gadde:2010en}. These matchings are highly non-trivial: the indices turn out as integrals of elliptic hypergeometric functions, which by themselves form an active field of mathematical research \cite{spiridonov2001elliptic,spiridonov2004theta,spiridonov2008essays,rains2003transformations}. It turned out that the identities that were needed in order to show equality of the indices of dual SQCD's, for example, were proved by mathematicians only a few years before they were obtained by Dolan and Osborn as the condition for the matching of the indices \cite{Dolan:2008qi}.
The R\"omelsberger index therefore provides a new, powerful, test for duality.

Motivated by the remarkable matching between indices of dual theories, followed a series of suggestions of new dualities and new mathematical identities, based on calculations of indices \cite{Spiridonov:2008zr,Spiridonov:2009za,Spiridonov:2010hh,Spiridonov:2011hf}.
The superconformal index technique was also applied to test conjectured dualities in theories with extended supersymmetry \cite{Gadde:2009kb,Spiridonov:2010qv}.

The highly non-trivial agreement between indices of dual theories also supports the claim that the index is invariant under the RG flow. This is because the duality tests used indices that were computed using the free theory, even for theories that are not IR free. It is therefore widely believed that the R\"omelsberger index is an RG-invariant.

Yet, as Witten demonstrated in \cite{Witten:1982df}, couplings that govern the large field behavior of the scalar potential can cause singularities that destroy the invariance of the index. Our first goal in this work is to explain why, even though couplings that are dominant in the large field limit may appear in the interacting theory, they cannot cause any singularity that threaten the invariance of the R\"omelsberger index.

To define the R\"omelsberger index one needs to choose a supercharge $Q$. The index counts zeros of $\{Q,Q^\dagger\}$, weighted by their eigenvalues under a set of operators, which commute with $Q$ and with each other, and accompanied by a sign, which is negative for fermionic states and positive for bosonic states.

In this work, we studied a free theory of chiral multiplets on $S^3\times{\mathbb{R}}$ by expanding the fields in $S^3$ spherical harmonics, and we found that each term in the decomposition of the bosonic part of $\{Q,Q^\dagger\}$ is equivalent to a Landau Hamiltonian (in the symmetric gauge) describing an electron in a constant magnetic field. The role of the magnetic field is played by the inverse of the radius of $S^3$. The eigenfunctions of $\{Q,Q^\dagger\}$ in field space are the same as the eigenfunctions of the Landau Hamiltonian in position space, and therefore they decay exponentially with the magnitude of the scalar fields.
We used this result to exclude a singular disappearance of low states when a superpotential is switched on: since the eigenfunctions whose eigenvalues we count are concentrated in a region in field space in which the superpotential is small, states of zero $\{Q,Q^\dagger\}$ may gain a positive eigenvalue, but only in a way which is continuous in the coupling constants. The index is invariant under such continuous changes.
The full argument to exclude these singularities appears in section \ref{argument_why}. The same argument shows that singularities in the scalar potential do not appear also when a gauge coupling is switched on.

In the second part of this work we prove a constraint on the RG flow of asymptotically free theories of chiral and vector multiplets, with unbroken supersymmetry and a non-anomalous R-symmetry.\\*
{\it Assuming:}
\begin{itemize}
\item{The R-charges of the chiral superfields of the UV theory are $q_i\in(0,2)$}.
\item{The IR theory is a free theory of chiral bound states with R-charges $\tilde{q}_j$}.
\end{itemize}
{\it then:}
\begin{itemize}
\item{$\tilde{q}_j\in(0,2)$.}
\end{itemize}

  When applied to theories that contain only one anomaly free R symmetry, this constraint is a weak version of a conjecture suggested by Intriligator. In \cite{Intriligator:2005if} Intriligator proposed a criterion for determining the IR phase of 4d quantum field theories, according to which the correct IR phase is the one with the larger conformal anomaly $a$. He also suggested a stronger conjecture: ``an operator can become IR free only if that results in a larger conformal anomaly $a$.'' For supersymmetric theories $a$ is given in terms of the superconformal R-charges, which are determined using $a$-maximization \cite{Intriligator:2003jj}. An operator with R charge $>{5/3}$ will contribute to $a$ more than a free operator with R-charge ${2/3}$ and therefore, for supersymmetric theories, Intriligator's stronger conjecture translates into: ``operators with R-charge $>{5/3}$ do not become IR free'' \cite{Intriligator:2005if}. This is the conjecture that is stronger than what we have proven. Note that the constraint we proved applies only to IR free theories, while Intriligator's conjecture applies also to theories where some operators become free and decouple, while the others interact between themselves.

Naively, the constraint above follows directly from the invariance of the index under the RG:
the contribution of a chiral multiplet to the index takes the form
\begin{equation}
I(t,y)=\prod_{m=0}^{\infty}\prod_{n=0}^{\infty}{{1-t^{m+n+2-q}y^{m-n}}\over{1-t^{m+n+q}y^{m-n}}}\;,\label{1}
\end{equation}
where $q$ is the R-charge of the chiral multiplet.
$I(t,y)$ includes negative powers of $t$ if $q>2$ and therefore invariance of the index under the RG flow seems to imply a constraint similar to the one suggested above. However, for (\ref{1}) to have a conventional interpretation as a generating function, with a good  formal expansion at $t=0$, we need to take $q\in(0,2]$. Since the interpretation of (\ref{1}) for $q\notin(0,2]$ is not currently clear to us, we prefer not to use the index in the argument proving the constraint.

Instead of using the full index, we used the fact that the difference between the number of bosonic and fermionic states that belong to the kernel of $\{Q,Q^\dagger\}$ and have a specific set of eigenvalues under our set of commuting operators does not depend on the coupling constants. In section \ref{application} we argue that if the constraint is not satisfied we get a contradiction with this rule.

The constraint above can be used to determine the correct IR phase for the model studied by Intriligator,
Seiberg and Shenker in \cite{Intriligator:1994rx}, which is an $SU(2)$ gauge theory with one chiral superfield $Q_{\alpha\beta\gamma}$ in the ${3/2}$ representation. This theory has one anomaly free R-symmetry, under which $Q$ has R-charge $3/5$. The basic gauge invariant operator in this theory is $u=Q_{\alpha\beta\gamma}Q^{\alpha\beta\rho}Q^{\sigma\lambda\gamma}Q_{\sigma\lambda\rho}$, with R-charge ${12/5}$. In \cite{Intriligator:1994rx} two possibilities were suggested for the IR phase of this theory near $u=0$. The first was a free theory of confined composites, with the only massless fields being the $u$ quanta, and the second was an interacting superconformal field theory. The first possibility was considered more likely to be the correct IR phase due to non-trivial 't Hooft anomaly matching. However, in \cite{Brodie:1998vv} it was demonstrated that non-trivial anomaly matchings are sometimes misleading. Since then, several works have suggested that the correct IR phase is actually the interacting superconformal theory \cite{Intriligator:2005if,Poppitz:2009kz,Vartanov:2010xj,Buican:2011ty}. According to our constraint this is indeed the case, since the IR free phase contains a bound state $u$ with R-charge ${12/5}>2$, and is therefore excluded.
The R\"omelsberger index was already used to determine the IR phase of this theory in \cite{Vartanov:2010xj}. We used a slightly different argument in our proof of the general constraint, since the physical interpretation of (\ref{1}) for $q>2$ is not clear to us.

The outline of the paper is as follows.
Section \ref{evaluations} starts with a theory of free chiral multiplets on $S^3\times\mathbb{R}$, which was constructed in \cite{Festuccia:2011ws}. We use an expansion in $S^3$ spherical harmonics to investigate this theory. The quantization of each mode of the theory is identical to the quantization of the Landau Hamiltonian in the symmetric gauge, and results in eigenstates that are localized around the origin of field space. We reproduce the R\"omelsberger index and use the localized eigenfunctions to claim that the singularities threatening the invariance of the index do not appear.
In section~\ref{application} we prove the constraint discussed above.
In appendix \ref{conventions} we summarize the conventions we use and in appendix \ref{spherical_harmonics} we collect some of the useful properties of the spherical harmonics on $S^3$. In appendix \ref{gauge theories} we couple gauge fields to the theory of section \ref{evaluations} and use the mode decomposition of this theory to reproduce the gauge multiplet contribution to the index.

\section{The Index on $S^3\times{\mathbb{R}}$}\label{evaluations}
\subsection{Short Review of SUSY on $S^3\times{\mathbb{R}}$ and the R\"omelsberger Index}

The isometry group of $S^3$ is $SU(2)_L\times{SU(2)_R}$. The supercharges $Q_\alpha$ sit in a doublet of $SU(2)_L$ and are singlets under $SU(2)_R$. The generators of the two $SU(2)$'s will be denoted by $J^i_L$ and $J^i_R$ ($i=1,2,3$). In addition, for the theory to conserve time-independent supercharges, it must conserve a $U(1)_R$ symmetry, with generator $R$ \cite{Festuccia:2011ws}.
The anticommutation relations of the supercharges are \cite{Festuccia:2011ws}
\begin{equation*}
\begin{aligned}
\{Q_\alpha,\bar{Q}_{\dot\alpha}\}&=-2\sigma^0_{\alpha\dot\alpha}(H-{1\over r}R)+{2\over r}\sigma^i_{\alpha\dot\alpha}J^i_L\;,\\
\{Q_\alpha,Q_\beta\}&=0\;,
\end{aligned}
\end{equation*}
where $H$ is the generator of translations along ${\mathbb{R}}$, and $r$ is the radius of $S^3$ (our conventions for the sigma matrices are summarized in appendix~\ref{conventions}).
In particular,
\begin{align}
&\{Q_1,Q_1^\dagger\}=2[H-{1\over r}R+{1\over r}J^3_L]\;,\\
&\{Q_2,Q_2^\dagger\}=2[H-{1\over r}R-{1\over r}J^3_L]\;.\label{Delta_2}
\end{align}

To define the R\"omelsberger index we choose the supercharges $Q_2$ and $Q_2^{\dagger}$, which commute with the symmetries $R+J^3_L$ and $J^3_R$. To regulate the index and to store more information in it, two fugacities $t$ and $y$, corresponding to the symmetries $R+J^3_L$ and $J^3_R$, are introduced into the index. These fugacities were recently identified as coordinates on the moduli space of complex structures on $S^3\times S^1$ \cite{Closset:2013vra}. The index is then defined as\footnote{We could have used an equivalent index: \begin{equation}I=\text{Tr}_{\Delta_1=0}\Big({(-1)^F}{t^{R-{J^3_L}}}{y^{J^3_R}}\Big)\;,\end{equation} where $\Delta_1={1\over2}\{Q_1,Q_1^{\dagger}\}$.  Since the supercharges form a doublet of $SU(2)_L$ the difference between $\Delta_1$ and $\Delta_2$ is manifested as $J^3_L \to -J^3_L$ and thus this index is identical to (\ref{index}).}
\begin{equation}
I(t,y)=\text{Tr}_{\Delta_2=0}\Big({(-1)^F}{t^{R+{J^3_L}}}{y^{J^3_R}}\Big)\;,\label{index}
\end{equation}
where $\Delta_2={1\over2}\{Q_2,Q_2^{\dagger}\}$ and $(-1)^F$ is defined to equal $1$ on bosonic states and $-1$ on fermionic states.

$I$ does not depend on the parameters of the theory, as long as the dependence of the low  $\Delta_2$ states and their eigenvalues on these parameters is continuous, and as long as the operators $R+J^3_L$ and $J^3_R$ do not depend on these parameters.
For example, $I$ is independent of a coupling constant if the dependence on this coupling is continuous, but it does depend on the R-charges of the chiral multiplets, since $R+J^3_L$ depends on these charges.

In \cite{Witten:1982df} Witten demonstrated that on $\mathbb{T}^4$, the set of low energy states may depend discontinuously on a coupling constant, thus allowing a discontinuous dependence of $\text{Tr}(-1)^F$ on this coupling. He considered the potential:
\begin{equation*}
V(\phi)=(m\phi-g\phi^2)^2\;.
\end{equation*}
For $g=0$ low energy states correspond to eigenfunctions $f(\phi)$ concentrated around $\phi=0$. For $g\neq0$ low energy eigenfunction $f(\phi)$ can be large also in a small region around $\phi={m/g}$. Since $m/g$  tends to infinity when $g$ tends to zero Witten referred to this situation as new low energy states ``moving in from infinity."
We wish to show that singularities of this type cannot appear on $S^3\times{\mathbb{R}}$.

We start by considering a supersymmetric theory of a free chiral superfield $\Phi=(\phi,\psi,F)$ with R-charge $q$ on $S^3\times{\mathbb{R}}$.
We next quote results of \cite{Festuccia:2011ws}, in which this theory was derived.

The supersymmetric lagrangian describing a free chiral superfield on $S^3\times{\mathbb{R}}$ is given by:
\begin{equation}
\begin{aligned}
&{{\cal L}^B}_{free}=|\partial_0\phi|^2-i{{q-1}\over{r}}(\phi\partial_0\bar\phi-\bar\phi\partial_0\phi)+{{q(q-2)}\over{r^2}}|\phi|^2-|\partial_i\phi|^2+|F|^2\;,\\
&{{\cal L}^F}_{free}=-i\bar\psi\bar\sigma^0\left(\partial_0-i{{q-{1\over{2}}}\over{r}}\right)\psi-i\bar\psi\bar\sigma^i\nabla_i\psi\;,
\end{aligned}
\label{Lagrangian-free-chiral}
\end{equation}
where $\nabla$ represents the covariant derivative on the sphere.

The SUSY variations of the fields are:
\begin{equation}
\begin{aligned}
&\delta\phi\;=-\sqrt2\xi\psi\;,\\
&\delta\psi_\alpha=-\sqrt2\xi_\alpha{F}-i\sqrt2[(\sigma^0\bar\xi)_\alpha\left(\partial_0-i{q\over{r}}\right)+(\sigma^i\bar\xi)_\alpha\partial_i]\phi\;,\\
&\delta{F}\;=-i\sqrt2\bar\xi[\bar\sigma^0\left(\partial_0-i{{q-{1\over{2}}}\over{r}}\right)+\bar\sigma^i\nabla_i]\psi\;.\\
\end{aligned}\label{trans-free-chiral}
\end{equation}

The action defined by (\ref{Lagrangian-free-chiral}) is invariant under the transformations (\ref{trans-free-chiral}) as long as the SUSY spinor satisfies:
\begin{equation}
\begin{aligned}
&\nabla_i\xi^{\alpha}={i\over{2r}}(\xi\sigma^i\bar\sigma^0)^{\alpha}\;,\\
&\partial_0\xi=0\;.
\end{aligned}
\label{spinor-eq-2}
\end{equation}
Equation (\ref{spinor-eq-2}) is a special case of Killing spinor equation discussed in \cite{Festuccia:2011ws}.

\subsection{Mode Expansion}\label{mode-expansion}
In this section we study the mode decomposition of the theory, which is obtained by expanding the fields in $S^3$ spherical harmonics. The properties of these functions and the conventions we use for them are summarized in Appendix~\ref{spherical_harmonics}.

We expand $\phi$ and $F$ in scalar spherical harmonics $\{\phi_{k,m,n}\}$, and $\psi$ in spinor spherical harmonics $\{\psi^\epsilon_{k,m,n}\}$:
\begin{equation}
\begin{aligned}
\phi&=\sum{a_{k,m,n}(t)\phi_{k,m,n}}(\theta,\alpha,\beta)\;,\\
\psi&=\sum{c^1_{k,m,n}(t)\psi^1_{k,m,n}(\theta,\alpha,\beta)}+\sum{c^{-1}_{k,m,n}(t)\psi^{-1}_{k,m,n}(\theta,\alpha,\beta)}\;,\\
F&=\sum{b_{k,m,n}(t)\phi_{k,m,n}(\theta,\alpha,\beta)}\;.
\end{aligned}
\label{modes-chiral}
\end{equation}
The mode decomposition of the Lagrangian (\ref{Lagrangian-free-chiral}) is given by:
\begin{equation*}
\begin{aligned}
{L}^B_{free}&=\sum_{k,m,n}[|\partial_0a_{k,m,n}|^2-i{{q-1}\over{r}}(a_{k,m,n}\partial_0a^*_{k,m,n}-a^*_{k,m,n}\partial_0a_{k,m,n})+\left({{q(q-2)}\over{r^2}}-{{k(k+2)}\over{r^2}}\right)|a_{k,m,n}|^2+|b_{k,m,n}|^2]\;,\\
{L}^F_{free}&=i\sum_{k,m,n}c^{1*}_{k,m,n}\left(\partial_0-i{{q-k-2}\over{r}}\right)c^1_{k,m,n}+i\sum_{k,m,n}c^{-1*}_{k,m,n}\left(\partial_0-i{{q+k}\over{r}}\right)c^{-1}_{k,m,n}\;.
\end{aligned}
\end{equation*}

We also expand the SUSY spinor $\xi$ in spinor spherical harmonics. $\xi$ is constrained by equation (\ref{spinor-eq-2}) such that only 2 modes survive:
\begin{equation}
\xi=\sqrt2\pi{r^{3\over2}}[\xi^1\psi^1_{0,1,0}(\theta,\alpha,\beta)+\xi^2\psi^1_{0,-1,0}(\theta,\alpha,\beta)]=\xi^1\left(\begin{array}{c}{1}\\{0}\end{array}\right)+\xi^2\left(\begin{array}{c}{0}\\{1}\end{array}\right)\;,
\end{equation}
where $\xi^1$ and $\xi^2$ are constant grassman numbers.

By setting $\xi^1=0$ in the Noether procedure we pick the supercharge $Q_2$. As we are interested in the the eigenfunctions and zero modes of the operator $\{ Q_2, {Q_2^\dagger}\}$, we write this operator as a sum of decoupled positive-semidefinite operators:
\begin{equation*}\{Q_2,{Q_2^\dagger}\}=2\sum_{k,m,n}(\Delta^B_{k,m,n}+\Delta^F_{k,m,n})\;,\end{equation*}
where:\footnote{The conjugate modes are:
\begin{equation*}
\Pi_{a_{k,m,n}}=(\partial_0+{i(q-1)\over r}){a^*_{k,m,n}},\;\;\;\Pi_{a_{k,m,n}^*}=(\partial_0-{i(q-1)\over r}){a_{k,m,n}},\;\;\;\Pi_{c^1_{k,m,n}}=-ic^{1*}_{k,m,n},\;\;\;\Pi_{c^{-1}_{k,m,n}}=-ic^{-1*}_{k,m,n}\;,
\end{equation*}
with the non-vanishing commutators and anticommutators being:
\begin{equation*}
{[}a_{k,m,n},\Pi_{a_{k,m,n}}{]}=i,\;\;\;
[a^*_{k,m,n},\Pi_{a^*_{k,m,n}}]=i,\;\;\;
\{c^1_{k,m,n},\Pi_{c^1_{k,m,n}}\}=-i,\;\;\;
\{c^{-1}_{k,m,n},\Pi_{c^{-1}_{k,m,n}}\}=-i\;.
\end{equation*}}\footnote{The sums in the expressions for $\Delta_2$ are over the indices:
\begin{equation*}
k\geq0\;\;\;,\;\;\;|m|,|n|\leq{k}\;\;\;,\;\;\;m \equiv n \equiv k \pmod 2\;.
\end{equation*}}

\begin{equation}
\begin{aligned}
\Delta^B_{k,m,n}&=|\Pi_{a_{k,m,n}}|^2+{{(1+k)^2}\over{r^2}}|a_{k,m,n}|^2+{i(m+1)\over r}(a^*_{k,m,n}\Pi_{a^*_{k,m,n}}-\Pi_{a_{k,m,n}}a_{k,m,n})-{{k-m}\over r}\;,\\
\Delta^F_{k,m,n}&=+{i(k+m+2)\over r}\Pi_{c^1_{k,m+1,n}}c^1_{k,m+1,n}-{i(k-m)\over r}\Pi_{c^{-1}_{k,m+1,n}}c^{-1}_{k,m+1,n}+{{k-m}\over r}\;.
\end{aligned}\label{delta_2}
\end{equation}

In the bosonic part we have operators of the form:
\begin{equation}
\Delta^B_{k,m}=|\Pi_a|^2+{{(1+k)^2}\over{r^2}}|a|^2+{i(m+1)\over r}(a^*\Pi_{a^*}-\Pi_{a}a)-{{k-m}\over r}\;.\label{Delta_B}
\end{equation}
After switching to real variables
 \begin{equation*}
a={1\over{\sqrt2}}(a_1+ia_2),\;\;\;
\Pi_a={1\over{\sqrt2}}(\Pi_1-i\Pi_2)\;,
\end{equation*}
and denoting
\begin{equation*}
L_3=a_1\Pi_2-a_2\Pi_1,\;\;\;\omega_k={2\over r}(1+k)\;,
\end{equation*}
we obtain:
\begin{equation*}
\Delta^B_{k,m}={1\over 2}(\Pi_1^2+\Pi_2^2)+{\omega_k^2\over 8}(a_1^2+a_2^2)-{\omega_k\over 2}L_3-{\omega_k\over 2}+{{k-m}\over r}L_3\;.
\end{equation*}

For $k=m$ modes, $\Delta^B_{k,k}$ is the Hamiltonian of an electron in a constant magnetic field, with the Landau levels shifted by $-{\omega_k\over 2}$ such that the lowest energy is zero (more precisely, the Hamiltonian for an electron in a constant magnetic field assumes this form in the symmetric gauge).
The eigenfunctions of this operator are most easily expressed in terms of polar coordinates $(\rho,\varphi)$ as:
\begin{equation}
\psi_{\tilde{n},\tilde{m}}(\rho,\varphi)=Ce^{-{\omega_k\over 4}\rho^2}{\rho^{|\tilde{m}|}}L_{\tilde{n}}^{|\tilde{m}|}\left({\omega_k\over{2}}\rho^2\right)e^{i\tilde{m}\varphi}\;,
\label{eigenfunctions}\end{equation}
where $L_{\tilde{n}}^{|\tilde{m}|}$ is a generalized Laguerre polynomial, $\tilde{m}$ is an integer, $\tilde{n}$ a non-negative integer, and $C$ is a constant.
For $k\neq m$ modes, the additional term ${{k-m}\over r}L_3$ does not change the eigenfunctions and only shifts the eigenvalues.\footnote{The eigenvalues of $\Delta^B_{k,m,n}$ corresponding to the eigenfunctions above are: \begin{equation}
\omega_k{l}+{{k-m}\over r}\tilde{m}\;,\label{spectrum}
\end{equation}
with $l=\tilde{n}+{1\over 2}(|\tilde{m}|-\tilde{m})\geq{max}(0,-\tilde{m})$.}

The $e^{-{\omega_k\over 4}\rho^2}=e^{{-{1\over {2r}}(1+k)}|a|^2}$ factor in (\ref{eigenfunctions}) localizes the eigenstates near the origin of field space. In subsection \ref{argument_why} we will use this fact to claim that low $\Delta$ states cannot come from or disappear into infinity when the superpotential is switched on.
The region to which $\psi_{\tilde{n},\tilde{m}}(\rho,\varphi)$ is localized depends on $|\tilde{m}|$ and $\tilde{n}$. Nevertheless, the eigenfunctions can be made as localized as we want by taking the radius of the three-sphere small enough. This means that whenever we are dealing with an object that gets contributions from a finite number of eigenfunctions we will be able to localize them as much as we want by choosing $r$ small enough. This will serve us in section \ref{argument_why} to claim that the first $N$ terms in the index are protected from the singularity. For larger $N$ we will have to take smaller radius for $S^3$, but since the index does not depend on the radius of the sphere the claim will follow for all the terms in the index.

To calculate the index, it is convenient to diagonalize $\Delta^B_{k,m,n}$ using the ladder operators:
\begin{align*}
A^1_{k,m,n}&=\sqrt{r\over 2(1+k)}\left(\Pi_{a_{k,m,n}}-i{{1+k\over r}}a^*_{k,m,n}\right)\;,\\
A^{1\dagger}_{k,m,n}&=\sqrt{r\over 2(1+k)}\left(\Pi_{a^*_{k,m,n}}+i{{1+k\over r}}a_{k,m,n}\right)\;,\\
A^2_{k,m,n}&=\sqrt{r\over 2(1+k)}\left(\Pi_{a^*_{k,m,n}}-i{{1+k\over r}}a_{k,m,n}\right)\;,\\
A^{2\dagger}_{k,m,n}&=\sqrt{r\over 2(1+k)}\left(\Pi_{a_{k,m,n}}+i{{1+k\over r}}a^*_{k,m,n}\right)\;,
\end{align*}
in terms of which we obtain:\footnote{According to (\ref{DeltaB}) the spectrum of  $\Delta^B_{k,m,n}$ is:
$N_{A1}{{k-m}\over r}+N_{A2}{{k+m+2}\over r}$, with $N_{A1},N_{A2}$ non negative integers.
By defining $l=N_{A2},\;\;\tilde{m}=N_{A1}-N_{A2}$, we get (\ref{spectrum}) back with $l,\tilde{m}$ integers and $l\geq{max}(0,-\tilde{m})$, as before.}
\begin{equation}
\Delta^B_{k,m,n}={{k-m}\over r}A^{1\dagger}_{k,m,n}A^1_{k,m,n}+{{k+m+2}\over r}A^{2\dagger}_{k,m,n}A^2_{k,m,n}\;.
\label{DeltaB}\end{equation}

Using the fermionic creation and annihilation operators
\begin{align*}
B^1_{k,m,n}=c^1_{k,m,n}\;\;,\;\;B^{1\dagger}_{k,m,n}=c^{1*}_{k,m,n}\;\;,\;\;B^2_{k,m,n}=c^{-1*}_{k,m,n}\;\;,\;\;B^{2\dagger}_{k,m,n}=c^{-1}_{k,m,n}\;,
\end{align*}
$\Delta^F_{k,m,n}$ is written as:
\begin{equation*}
\Delta^F_{k,m,n}={(k+m+2)\over r}{B^{1\dagger}_{k,m+1,n}}B^1_{k,m+1,n}+{(k-m)\over r}{B^{2\dagger}_{k,m+1,n}}B^2_{k,m+1,n}\;.
\end{equation*}

\subsection{Reproducing the Index}
We want to calculate the index:
\begin{equation}
I=\text{Tr}_{\Delta_2=0}\Big({(-1)^F}{t^{R+{J^3_L}}}{y^{J^3_R}}\Big)\;,\label{index1}
\end{equation}
where:
\begin{align*}
J^3_L&=\sum_{k,m,n}m(A^{1\dagger}_{k,m,n}A^1_{k,m,n}-A^{2\dagger}_{k,m,n}A^2_{k,m,n})-\sum_{k,m,n}m{B^{1\dagger}_{k,m,n}}B^1_{k,m,n}+\sum_{k,m,n}m{B^{2\dagger}_{k,m,n}}B^2_{k,m,n}\;,\\
J^3_R&=\sum_{k,m,n}n(A^{1\dagger}_{k,m,n}A^1_{k,m,n}-A^{2\dagger}_{k,m,n}A^2_{k,m,n})-\sum_{k,m,n}n{B^{1\dagger}_{k,m,n}}B^1_{k,m,n}+\sum_{k,m,n}n{B^{2\dagger}_{k,m,n}}B^2_{k,m,n}\;,\\
R&=\sum _{k,m,n}q(A^{1\dagger}_{k,m,n}A^1_{k,m,n}-A^{2\dagger}_{k,m,n}A^2_{k,m,n})-\sum _{k,m,n}(q-1)B^{1\dagger}_{k,m,n}B^1_{k,m,n}+\sum _{k,m,n}(q-1)B^{2\dagger}_{k,m,n}B^2_{k,m,n}\;,
\end{align*}
and
\begin{equation}
\begin{aligned}
\Delta_2=\sum_{k,m,n}\Big(&{{k-m}\over r}A^{1\dagger}_{k,m,n}A^1_{k,m,n}+{{k+m+2}\over r}A^{2\dagger}_{k,m,n}A^2_{k,m,n}\\
&+{(k+m+2)\over r}{B^{1\dagger}_{k,m+1,n}}B^1_{k,m+1,n}+{(k-m)\over r}{B^{2\dagger}_{k,m+1,n}}B^2_{k,m+1,n}\Big)\;.
\end{aligned}\label{d2}
\end{equation}

The kernel of $\Delta_2$ is generated by the subset of the creation operators composed of $A^{1\dagger}_{k,k,n}$ and $B^{1\dagger}_{k,-k-1,n}$.\footnote{It is implicit in (\ref{d2}) that the sum is over indices corresponding to an expansion in scalar spherical harmonics, therefore $B^1_{k,-k-1,n}$ do not appear in (\ref{d2}) and are zero modes. On the other hand, the coefficient of $B^{2\dagger}_{k,m+1,n}B^2_{k,m+1,n}$ vanishes for $k=m$, but the $\psi^{-1}_{k,m,n}$ harmonics only go up $m=k-1$, and therefore there are no $B^2$ zero modes.}

For $q\leq0$ there are infinitely many $\Delta_2=0$ states with $R+J^3_L=J^3_R=0$ and thus the sum in (\ref{index1}) is ill defined. We do not know if this divergence can be fixed by an appropriate regulator. In the following we will assume that $q>0$.

The index is calculated using a combinatorial trick \cite{Benvenuti:2006qr,Feng:2007ur,Kinney:2005ej,Romelsberger:2007ec}: first the ``single particle states" contributions are summed to a generating function and then the index is calculated as the plethystic exponential of the generating function.
The bosonic single particle zero modes have the following eigenvalues:
\begin{equation}
\begin{aligned}
&R+J^3_L=q+k\;,\\
&J^3_R=n\;.
\end{aligned}\label{eigenvalues-b}
\end{equation}
Thus the bosonic contribution to the generating function is:
\begin{equation*}
\sum_{k}t^{q+k}\sum_{n}y^{n}={t^q\over {(1-{t\over y})(1-ty)}}\;.
\end{equation*}

The fermionic single particle zero modes have the eigenvalues:
\begin{equation}
\begin{aligned}
&R+J^3_L=-(q-1)-(-k-1)=k+2-q\;,\\
&J^3_R=-n\;.
\end{aligned}\label{eigenvalues-f}
\end{equation}
and thus the fermionic generating function is:
\begin{equation*}
-\sum_{k}t^{k+2-q}\sum_{n}y^{-n}=-{t^{2-q}\over {(1-{t\over y})(1-ty)}}\;.
\end{equation*}

The complete index is:
\begin{equation}
\begin{aligned}
I&=\text{exp}\left(\sum_{l=1}^{\infty }{1\over l}(\sum_{k,n}t^{l(q+k)}y^{ln}-\sum_{k,n}t^{l(k
+2-q)}y^{-ln}) \right)\\&=\prod_{k,n}{{1-t^{k
+2-q}y^n}\over{1-t^{k
+q}y^n}}=\prod_{m'=0}^{\infty}\prod_{n'=0}^{\infty}{{1-t^{m'+n'
+2-q}y^{m'-n'}}\over{1-t^{m'+n'
+q}y^{m'-n'}}}=\Gamma(t^q;ty,t/y)\;,
\end{aligned}\label{index-chiral}
\end{equation}
where $\Gamma$ is an elliptic gamma function.
This result agrees with the index calculated by R\"omelsberger in \cite{Romelsberger:2007ec}.

In order for the index to have a conventional interpretation as a generating function, with a good  formal expansion at $t=0$, we need to take $q\in(0,2]$.

\subsection{Why no Vacua can Come from Infinity}\label{argument_why}
In order to obtain better understanding of the way the presence of the superpotential affects the low $\Delta$ states, we start by considering the QM theory obtained by truncating the mode expansion to include only the $k=0$ modes, and add superpotential interactions:
\begin{align*}
{L}_{k=0}=&|F|^2+|\partial_{0}a|^2+{i(1-q)\over r}\left(a\partial_{0}a^*-a^*\partial_{0}a\right)+{q(q-2)\over r^2}|a|^2+ic_{-1}^*\left(\partial_{0}-i{{q-2}\over r}\right)c_{-1}\\
&+ic_1^*\left(\partial_{0}-i{{q-2}\over r}\right)c_1
-[c_1c_{-1}W''(a)+FW'(a)+c.c.]\;,
\end{align*}
where the following abbreviated notations have been used:
\begin{equation*}
a={a_{0,0,0}}\;\;\;,\;\;\;c_1={c^1_{0,1,0}}\;\;\;,\;\;\;c_{-1}={c^1_{0,-1,0}}\;\;\;,\;\;\;F={b_{0,0,0}}\;,
\end{equation*}
and $W(a)$ is of the form
\begin{equation*}
W(a)=\lambda{a}^n\;\;\;,\;\;\;n={2\over q}\;\;\;,\;\;\;\lambda \in \mathbb{C}\;.
\end{equation*}

In the presence of this superpotential, the lowest modes truncation of (\ref{Delta_B}) gets an extra contribution of
\begin{equation*}
|W'(a)|^2=|\lambda|^2n^2(|a|^2)^{n-1}\;,
\end{equation*}
and becomes
\begin{equation*}
\Delta_2^B=|\Pi_a|^2+{1\over r^2}|a|^2+{i\over r}\left(a^*\Pi_{a^*}-\Pi_aa\right)+|\lambda|^2n^2(|a|^2)^{n-1}\;.
\end{equation*}

This expression explains why low $\Delta$ states do not suddenly appear or disappear when the superpotential is turned on: new low $\Delta$ states cannot suddenly appear
since the additional term is non-negative and therefore including it can only constrain further the low lying states. Low $\Delta$ states do not suddenly disappear since the term proportional to $|a|^2$, which appears already in the free theory, localizes the low $\Delta$ states around $a=0$ (as can be seen explicitly in (\ref{eigenfunctions}));  For $0<q<2$ the additional perturbation is small in this region, and therefore low $\Delta$ states remain low $\Delta$ states - the additional term can raise the expectation value of $\Delta$, but this effect must be continuous in $|\lambda|^2$.
For $q=2$ the perturbation that is added is a SUSY breaking constant $|\lambda|^2$, in the presence of which no $\Delta=0$ states exist, and the index is zero. This term does not cause any singularity since the $\Delta$ eigenvalue that the states obtain is just $|\lambda|^2$ (therefore the continuity in $|\lambda|^2$ is trivial here). Indeed, it is easy to see that the index calculated for the free theory (\ref{index-chiral}) vanishes for $q=2$, even though the kernel of $\Delta$ is non-trivial.

The same argument holds also for many chiral multiplets with interactions between them. In this case the Lagrangian is of the form
\begin{align*}
{L}_{k=0}=&\sum_i[|F_i|^2+|\partial_{0}a_i|^2+{i(1-q_i)\over r}\left(a_i\partial_{0}a^{*}_i-a^{*}_i\partial_{0}a_i\right)+{q_i(q_i-2)\over r^2}|a_i|^2+ic_{-1i}^{*}\left(\partial_{0}-i{{q_i-2}\over r}\right)c_{-1i}\\
&+ic_{1i}^*\left(\partial_{0}-i{{q_i-2}\over r}\right)c_{1i}-\left(F_i{{\partial{W}}\over{\partial{a_i}}}+c.c.\right)]-\sum_{i,j}[c_{1i}c_{-1j}{{\partial^2W}\over{\partial{a_i}\partial{a_j}}}+c.c.]
\end{align*}
where $W(a_1,...,a_N)$ has R-charge 2,
and
\begin{equation}
\Delta_2^B=\sum_i[|\Pi_{a_i}|^2+{1\over r^2}|a_i|^2+{i\over r}\left(a_i^*\Pi_{a_i^*}-\Pi_{a_i}a_i\right)+|{{\partial{W}}\over{\partial{a_i}}}|^2]\;.\label{delta_2_b}
\end{equation}
Again, the superpotential contributions $|{{\partial{W}}\over{\partial{a_i}}}|^2$ are non-negative and bounded near the origin of field space.
By taking $r\to0$, we can make the second term in (\ref{delta_2_b}) dominant, thereby localizing all the wave functions near the origin.

Motivated by the lowest modes analysis we return to the full theory. As we discussed in subsection \ref{mode-expansion}, the eigenfunctions are localized in a small region around the origin of field space in which the perturbation is small, thus low $\Delta$ states do not disappear when the superpotential is switched on. Since the perturbation is by definition non-negative, new low $\Delta$ states cannot suddenly appear. We therefore expect the argument above to be valid in the full theory and the independence of the index in the superpotential couplings to be protected from the singularities discussed above.

In the following we will consider also gauge theories. The same argument shows that the gauge interactions contributions to the scalar potential do not cause this type of singularity.

\section{Application for the IR R-charges}\label{application}

In this section we prove a constraint on the RG flow of asymptotically free theories of chiral and vector multiplets, with unbroken supersymmetry and a non-anomalous R-symmetry.\\*
{\it Assuming:}
\begin{itemize}
\item{The R-charges of the chiral superfields of the UV theory are $q_i\in(0,2)$}.
\item{The IR theory is a free theory of chiral bound states with R-charges $\tilde{q}_j$}.
\end{itemize}
{\it then:}
\begin{itemize}
\item{$\tilde{q}_j\in(0,2)$.}\footnote{The superconformal R-charges of the free superfields are $2/3$, due to mixing with accidental symmetries. Here we do not refer to the superconformal R-charges, but to the charges of the composite operators with respect to the same R-symmetry we considered in the UV. Note also that this constraint applies to any anomaly free R-symmetry.}
\end{itemize}

This constraint provides a proof for a weak version of a conjecture proposed by Intriligator in \cite{Intriligator:2005if}.

As discussed in the introduction, we will not use the index in the argument below, since its interpretation for $q\notin(0,2]$ is not currently clear to us. Still, the proof is based on the same arguments used to claim that the index is topological.

We start with a free theory (the UV free fixed point), in which the chiral multiplets $\Phi_i$ have R-charges $q_i\in(0,2)$, and consider the spectrum of the operators $R+J^3_L$ and $J^3_R$, restricted to the kernel of $\Delta_2$.  Equations (\ref{eigenvalues-b}) and (\ref{eigenvalues-f}) show that $R+J^3_L$ does not have any negative eigenvalues in the kernel of $\Delta_2$ (the gauge sector does not contain negative eigenvalues of $R+J^3_L$, as can be seen in appendix~\ref{gauge theories} - this fact is of course independent on the R-charges of the chiral multiplets). The operators $R+J^3_L$ and $J^3_R$ do not depend on the coupling constants, and have a discrete spectrum. Thus, the difference between the number of bosonic and fermionic eigenfunctions that belong to the kernel of $\Delta_2$ and have specific eigenvalues of $R+J^3_L$ and $J^3_R$ does not depend on the coupling constants. Singularities such as those suggested by Witten in \cite{Witten:1982df}, whose existence could have threatened the independence of these differences in the coupling constants, were excluded by our argument in section \ref{argument_why}. We conclude that also in the IR theory - if the kernel of $\Delta_2$ contains states with negative eigenvalues under $R+J^3_L$, they come in degenerate pairs of boson and fermion.

 It is assumed that the IR theory also has a description in terms of a free theory of chiral bound states. If some of these bound states have R-charges larger than 2, (\ref{eigenvalues-b},\ref{eigenvalues-f}) tell us that the kernel of $\Delta_2$ contains a finite number of fermionic single particle states with negative eigenvalues of $R+J^3_L$, while the bosonic single particle states have only positive eigenvalues. By taking the product of all the single particle states with negative eigenvalue of $R+J^3_L$, we build an eigenstate $\left|\text{min} \right \rangle$ corresponding to the lowest eigenvalue that $R+J^3_L$ assumes in the kernel of $\Delta_2$. Since all the states with negative eigenvalues of $R+J^3_L$ must be arranged in degenerate pairs of boson and fermion, we conclude that $\left|\text{min} \right \rangle$ must have a partner with the same eigenvalues. This is possible only if the kernel of $\Delta_2$ contains a fermionic state $\left|0 \right \rangle$ with $R+J^3_L=J^3_R=0$ that is built only from fermionic single particle states with $R+J^3_L=0$ (since $\left|\text{min} \right \rangle$ already contains all the single particle states with negative eigenvalues). Looking again at (\ref{eigenvalues-f}) we see that the existence of $\left|0 \right \rangle$ must also imply the existence of a fermionic single particle state with $R+J^3_L=J^3_R=0$. But if the kernel of $\Delta_2$ contains a fermionic single particle state with $R+J^3_L=J^3_R=0$ then this state defines an invertible map between bosons and fermions with the same eigenvalues, which means that for every possible set of eigenvalues the difference between the number of bosonic and fermionic eigenstates is zero. Since not all of these differences vanish in the UV theory (for the UV theory with all the R-charges in $(0,2)$ we can simply use the calculated index, which is non-vanishing) we get a contradiction. We conclude that the IR theory cannot contain a chiral multiplet with R-charge larger than 2.

A chiral multiplet with R-charge 2 is also forbidden: here we can use the index, which is well-defined for $q\in(0,2]$. The index will vanish if the theory contains a chiral multiplet of R-charge 2, in contradiction to the fact that the index of the UV theory is non-vanishing. We therefore conclude that all the bound states in the IR theory must have their R-charges in the range $(0,2)$.

Our result is generalized for theories in which the IR free phase contains also gauge multiplets in the following way: under the same assumptions regarding the UV theory and assuming that the IR theory is described by free chiral and gauge multiplets, if the IR phase contains bound states with R-charge $\geq2$, then not all of them are gauge invariant.\footnote{Similar arguments to those we have used can be applied to prove further constraints.  For example, it is not difficult to show that under the same assumptions, if the UV R-charges are $q_i\in(0,1)$, and the IR theory is a free theory of chiral bound states, then at least one of these bound states has R-charge $\leq1$. It will be interesting to know if this constraint has applications.}

\section*{Acknowledgments}
First and foremost, I would like to thank Zohar Komargodski for suggesting this work as an M.Sc project,
and for his help and guidance throughout.
I would also like to thank Guido Festuccia, Shlomo Razamat, Itamar Shamir, Kenneth Intriligator and Matthew Buican for very helpful discussions and comments.
EG is supported by the ERC STG grant number 335182 and by the Israel Science Foundation under grant number 884/11. EG would also like to thank the United States-Israel Binational Science Foundation (BSF) for support under grant number 2010/629. In addition, the research of EG is supported by the I-CORE Program of the Planning and Budgeting Committee and by the Israel Science Foundation under grant number 1937/12.   Any opinions, findings, and conclusions or recommendations expressed in this material are those of the author and do not necessarily reflect the views of the funding agencies.

\appendix

\section{Conventions}\label{conventions}

We use the conventions of \cite{Wess:1992cp}. The flat space metric is:
\begin{equation*}
\eta_{\mu\nu}=(-1,1,1,1)\;,
\end{equation*}
and the sigma matrices are given by:
\begin{equation*}\sigma^0=\begin{pmatrix}-1 & 0 \\ 0 &-1\end{pmatrix}\;,\qquad\sigma^1=\begin{pmatrix}0 & 1 \\ 1 &0\end{pmatrix}\;,\qquad \sigma^2=\begin{pmatrix}0 & - i  \\ i &0\end{pmatrix}\;,\qquad \sigma^3=\begin{pmatrix}1 & 0 \\ 0 &-1\end{pmatrix}\;,\end{equation*}
\begin{equation*}
\bar\sigma^0=\sigma^0\qquad,\qquad\bar\sigma^{1,2,3}=-\sigma^{1,2,3}\;.
\end{equation*}

We use the following embedding of $S^3$ into $\mathbb{R}^4$:
\begin{align*}
&x_1=r\cos\theta\cos\alpha\;,\\
&x_2=r\cos\theta\sin\alpha\;,\\
&x_3=r\sin\theta\cos\beta\;,\\
&x_4=r\sin\theta\sin\beta\;.
\end{align*}
The inherited metric on $S^3$ is:
\begin{equation*}
ds^2=r^2[d\theta^2+\sin^2\theta d\beta^2+\cos^2\theta d\alpha^2]\;.
\end{equation*}

We also used the following choice of orthonormal frame on $S^3$:
\begin{align*}
&e^1_\theta=r\sin(\alpha+\beta)\;,\;\;\;\;\;\;\;\;\;\;e^1_\alpha=-{1\over2}r\sin(2\theta)\cos(\alpha+\beta)\;,\;\;\;\;\;\;\;e^1_\beta={1\over2}r\sin(2\theta)\cos(\alpha+\beta)\;,\\
&e^2_\theta=-r\cos(\alpha+\beta)\;,\;\;\;\;\;\;e^2_\alpha=-{1\over2}r\sin(2\theta)\sin(\alpha+\beta)\;,\;\;\;\;\;\;\;e^2_\beta={1\over2}r\sin(2\theta)\sin(\alpha+\beta)\;,\\
&e^3_\theta=0\;,\;\;\;\;\;\;\;\;\;\;\;\;\;\;\;\;\;\;\;\;\;\;\;\;\;\;e^3_\alpha=r\cos^2\theta\;,\;\;\;\;\;\;\;\;\;\;\;\;\;\;\;\;\;\;\;\;\;\;\;\;\;\;\;\;\;\;\;\;e^3_\beta=r\sin^2\theta\;.
\end{align*}

\section{Spherical Harmonics on $S^3$}\label{spherical_harmonics}
In this appendix we collect some properties of spherical harmonics on $S^3$ that we used in our analysis, most of them were taken from similar appendices in \cite{Mussel:2009uw,Aharony:2005bq}, in which they appeared in slightly different notations.

The scalar eigenfunctions of the laplacian on $S^3$ form a complete set of functions $\{\phi_{k,m,n}\}$, where the labels $\{k,m,n\}$ are integer numbers satisfying:
\begin{equation*}
k\geq0\;\;\;,\;\;\;|m|,|n|\leq{k}\;\;\;,\;\;\;m \equiv n \equiv k \pmod 2\;.
\end{equation*}
$\{\phi_{k,m,n}\}$ transform in the $({k\over2},{k\over2})$ representation of $SU(2)\times{SU(2)}$ and satisfy:
\begin{equation*}
\nabla^2\phi_{k,m,n}=-{1\over{r^2}}k(k+2)\phi_{k,m,n}\;.
\end{equation*}
They also obey the conjugation relation
\begin{equation*}
\phi^*_{k,m,n}=(-1)^{-{1\over{2}}(n+m)}\phi_{k,-m,-n}\;,
\end{equation*}
and the orthonormality relation\footnote{We are using a convention in which all the spherical harmonics (scalar spherical harmonics, vector spherical harmonics and spinor spherical harmonics) are multiplied by ${r^{-{3\over2}}}$ and have mass dimension ${3\over2}$.}
\begin{equation*}
\int_{S^3}^{}d{\theta}d{\alpha}d{\beta}\sqrt{g}\phi_{k,m,n}\phi^*_{k',m',n'}=\delta^{kk'}\delta^{mm'}\delta^{nn'}\;.
\end{equation*}
Spinors on the sphere can be expanded in a complete set of eigenfunctions of the Dirac operator, $\{\psi^\epsilon_{k,m,n}\}$, where  $\epsilon=\pm{1}$ and
\begin{equation*}
\begin{aligned}
k\geq0\;\;\;,\;\;\;|m|\leq{k+1}\;\;\;,\;\;\;|n|\leq{k}\;\;\;,\;\;\;m+1 \equiv n \equiv k \pmod 2\;\;\;&\text{for } \epsilon=1\;,\\
k\geq1\;\;\;;\;\;\;|m|\leq{k-1}\;\;\;,\;\;\;|n|\leq{k}\;\;\;,\;\;\;m+1 \equiv n \equiv k \pmod 2\;\;\;&\text{for } \epsilon=-1\;.
\end{aligned}
\end{equation*}
$\{\psi^1_{k,m,n}\}$ transform in the $({{k+1}\over2},{k\over2})$ representation of $SU(2)\times{SU(2)}$ and satisfy:
\begin{equation*}
\not{\nabla}\psi^{1}_{k,m,n}={i\over{r}}\left(k+{3\over{2}}\right)\psi^1_{k,m,n}\;,
\end{equation*}
where \begin{equation*}\not{\nabla}\psi=\sigma^0\bar\sigma^i\nabla_i\psi\;,\end{equation*}
while $\{\psi^{-1}_{k,m,n}\}$ transform in the $({{k-1}\over2},{k\over2})$ representation of $SU(2)\times{SU(2)}$ and satisfy:
\begin{equation*}
\not{\nabla}\psi^{-1}_{k,m,n}=-{i\over{r}}\left(k+{1\over{2}}\right)\psi^{-1}_{k,m,n}\;.
\end{equation*}
The spinor spherical harmonics can be expressed in terms of the scalar spherical harmonics:
\begin{equation*}
\begin{aligned}
&\psi^1_{k,m,n}={1\over{\sqrt{2(k+1)}}}\left(\begin{array}{c}{\sqrt{k+m+1}\phi_{k,m-1,n}}\\{\sqrt{k-m+1}\phi_{k,m+1,n}}\end{array}\right)\;,\\
&\psi^{-1}_{k,m,n}={1\over{\sqrt{2(k+1)}}}\left(\begin{array}{c}{\sqrt{k-m+1}\phi_{k,m-1,n}}\\{-\sqrt{k+m+1}\phi_{k,m+1,n}}\end{array}\right)\;.
\end{aligned}
\end{equation*}

Vector fields can be expanded in $\vec{V}^\epsilon_{k,m,n}$ and $\nabla\phi_{k,m,n}$.
$\vec{V}^\epsilon_{k,m,n}$ are the vector spherical harmonics.
They belong to the $({{k+2\epsilon}\over2},{k\over2})$ representation of $SU(2)\times{SU(2)}$, where $\epsilon=\pm{1}$. They are defined for
\begin{equation*}
\begin{aligned}
k\geq0\;\;\;,\;\;\;|m|\leq{k+2}\;\;\;,\;\;\;|n|\leq{k}\;\;\;,\;\;\;m \equiv n \equiv k \pmod 2\;\;\;&\text{for } \epsilon=+1\;,\\
k\geq2\;\;\;,\;\;\;|m|\leq{k-2}\;\;\;,\;\;\;|n|\leq{k}\;\;\;,\;\;\;m \equiv n \equiv k \pmod 2\;\;\;&\text{for } \epsilon=-1\;,
\end{aligned}
\end{equation*}
and they satisfy:
\begin{align*}
&\vec{\nabla}\cdot\vec{V}^\epsilon_{k,m,n}=0\;,\\
&\vec{\nabla}\times\vec{V}^\epsilon_{k,m,n}=-{\epsilon\over r}(k+1+\epsilon)\vec{V}^\epsilon_{k,m,n}\;,\\
&\nabla^2\vec{V}^\epsilon_{k,m,n}=-{1\over{r^2}}(k+1+\epsilon)^2\vec{V}^\epsilon_{k,m,n}\;.
\end{align*}

The vector spherical harmonics are orthonormal:
\begin{equation*}
\int_{S^3}{d{\theta}d{\alpha}d{\beta}\sqrt{g}\left(\vec{V}^\epsilon_{k,m,n}\cdot(\vec{V}^{{\epsilon}'}_{k',m',n'})^*\right)=\delta^{\epsilon{\epsilon}'}\delta^{kk'}\delta^{mm'}\delta^{nn'}}\;,
\end{equation*}
and obey the conjugation relation
\begin{equation}
(V^\epsilon_{k,m,n})^*=(-1)^{{{m+n}\over2}+1}V^\epsilon_{k,-m,-n}\;.\label{conjugation}
\end{equation}

In order to write the supercharges in terms of the modes we needed to calculate (\ref{integral}), which includes integral of a product of two spinor and one vector spherical harmonics. This integral was expressed in \cite{Mussel:2009uw} as a product of reduced matrix element and 3-j symbols:
\begin{equation*}
i{r^{3\over2}}\int_{S^3}{d\theta{d\alpha}d\beta}\sqrt{g}(\psi^{\epsilon_\alpha}_{k_\alpha,m_\alpha,n_\alpha})^\dagger\sigma_i\psi^{\epsilon_\beta}_{k_\beta,m_\beta,n_\beta}V^{\epsilon_\gamma,i}_{k_\gamma,m_\gamma,n_\gamma}=\end{equation*}
\begin{equation}
(-1)^{{{m_\alpha+n_\alpha+\epsilon_\alpha}\over2}}R(k_\alpha+{{1+\epsilon_\alpha}\over2},k_\beta+{{1+\epsilon_\beta}\over2},k_\gamma+\epsilon_\gamma;\epsilon_\alpha,\epsilon_\beta,\epsilon_\gamma)\begin{pmatrix}
   {{k_\alpha+\epsilon_\alpha}\over2} & {{k_\beta+\epsilon_\beta}\over2} & {{k_\gamma+2\epsilon_\gamma}\over2} \\
   -m_\alpha\over2 & m_\beta\over2 &  m_\gamma\over2  \\ \end{pmatrix}\begin{pmatrix}
   {k_\alpha}\over2 & {k_\beta}\over2 &  {k_\gamma}\over2 \\
   -n_\alpha\over2 & n_\beta\over2 &  n_\gamma\over2 \\
  \end{pmatrix}\;,\label{formula}
\end{equation}
where the reduced matrix elements are given by:
\begin{align*}
&R(x,y,z;1,1,1)=R(x,y,z;-1,-1,-1)={{(-1)^{\tilde\sigma+1}}\over\pi}\sqrt{{(\tilde\sigma-x)(\tilde\sigma-y)\tilde\sigma(\tilde\sigma+1)}\over{z+1}}\;,\\
&R(x,y,z;-1,-1,1)=-R(x,y,z;1,1,-1)={{(-1)^{\tilde\sigma+1}}\over\pi}\sqrt{{(\tilde\sigma-x)(\tilde\sigma-y)(\tilde\sigma-z-1)(\tilde\sigma-z)}\over{z+1}}\;,\\
&R(x,y,z;1,-1,1)=R(x,y,z;-1,1,-1)={{(-1)^{\sigma}}\over\pi}\sqrt{{(\sigma-z)(\sigma+1)(\sigma-y)(\sigma-y+1)}\over{z+1}}\;,\\
&R(x,y,z;-1,1,1)=R(x,y,z;1,-1,-1)={{(-1)^{\sigma+1}}\over\pi}\sqrt{{(\sigma-z)(\sigma+1)(\sigma-x)(\sigma-x+1)}\over{z+1}}\;,
\end{align*}
and:
\begin{equation*}
\sigma\equiv{1\over2}(x+y+z)\;\;\;,\;\;\;\tilde\sigma\equiv{1\over2}(x+y+z+1)\;.
\end{equation*}

\section{Gauge Theories on $S^3\times\mathbb{R}$}\label{gauge theories}
For simplicity we take the gauge group to be $U(1)$.

The Lagrangian describing free $U(1)$ gauge multiplet in the Wess-Zumino gauge $(A_\mu,\lambda_\alpha,D)$ coupled to free chiral multiplet $(\phi,\psi_\alpha,F)$ on $S^3\times\mathbb{R}$ can be extracted from \cite{Sohnius:1982fw}:
\begin{equation*}
\begin{aligned}
{{\cal L}}_{gauge}=&-{1\over 4}F_{\mu\nu}F^{\mu\nu}-i\bar\lambda\bar\sigma^0\left(\partial_0+{i\over {2r}}\right)\lambda-i\bar\lambda\bar\sigma^i\nabla_i\lambda+{1\over 2}D^2\;,\\
{{\cal L}}_{free}\;=&|(\partial_0-igA_0)\phi|^2-i{{q-1}\over{r}}\left(\phi(\partial_0+igA_0)\bar\phi-\bar\phi(\partial_0-igA_0)\phi\right)+{{q(q-2)}\over{r^2}}|\phi|^2-|(\partial_i-igA_i)\phi|^2+|F|^2\\
&-i\bar\psi\bar\sigma^0\left(\partial_0-igA_0-i{{q-{1\over{2}}}\over{r}}\right)\psi-i\bar\psi\bar\sigma^i(\nabla_i-igA_i)\psi-i{\sqrt2}g(\bar{\phi}\psi\lambda-\bar\lambda\phi\bar{\psi})-gD\bar{\phi}{\phi}\;,
\end{aligned}
\end{equation*}
where $F_{\mu\nu}=\partial_{\mu}A_\nu-\partial_{\nu}A_\mu$. As before $q$ is the R-charge of the chiral multiplet, $r$ is the radius of the three-sphere, and $\nabla$ is the covariant derivative on the sphere.

For $\xi$ satisfying (\ref{spinor-eq-2}), the action defined by these Lagrangians is invariant under the variations:
\begin{equation}
\begin{aligned}
&\delta\phi\;=-\sqrt2\xi\psi\;,\\
&\delta\psi_\alpha=-\sqrt2\xi_\alpha{F}-i\sqrt2(\sigma^\mu\bar\xi)_{\alpha}{D_\mu}\phi\;,\\
&\delta{F}\;=-i\sqrt2\bar\xi[\bar\sigma^\mu{D}_\mu\psi-{i\over{2r}}\bar\sigma^0\psi]+2ig\phi\bar\xi\bar\lambda\;,\\
&\delta{A_{\mu}}=-i\xi\sigma_\mu\bar\lambda+i\lambda\sigma_\mu\bar\xi\;,\\
&\delta\lambda_\alpha=-(\sigma^{\mu\nu}\xi)_{\alpha}F_{\mu\nu}-i\xi_\alpha{D}\;,\\
&\delta{D}\;=\xi\sigma^\mu{D_\mu}\bar\lambda+D_\mu\lambda\sigma^\mu\bar\xi-{3i\over{2r}}(\xi\sigma^0\bar\lambda-\lambda\sigma^0\bar\xi)\;,
\end{aligned}\label{var-s^3}
\end{equation}
where:
\begin{equation*}
\begin{aligned}
&D_{\mu}\phi\;\;=(\nabla_{\mu}-iq\tilde{A_{\mu}}-igA_{\mu})\phi\;,\\
&D_{\mu}\psi\;\;=(\nabla_{\mu}-i(q-1)\tilde{A_{\mu}}-igA_{\mu})\psi\;,\\
&D_{\mu}F\;\;=(\nabla_{\mu}-i(q-2)\tilde{A_{\mu}}-igA_{\mu})F\;,\\
&D_{\mu}A_{\nu}=\nabla_{\mu}{A_{\nu}}\;,\\
&D_{\mu}\lambda\;\;=(\nabla_{\mu}-i\tilde{A_{\mu}})\lambda\;,\\
&D_{\mu}D\;=\nabla_{\mu}D\;,\\
&D_{\mu}\xi\;\;=(\nabla_{\mu}-i\tilde{A_{\mu}})\xi\;,
\end{aligned}
\end{equation*}
with $\tilde{A_0}={1\over r},\;\;\;\tilde{A_i}=0$.\footnote{In the approach of \cite{Festuccia:2011ws} for placing the theory on $S^3\times\mathbb{R}$ by fixing the metric and auxiliary fields in the new minimal supergravity multiplet on a supersymmetry preserving values, $D_\mu$ represents a derivative that is covariant with respect to the geometry of the sphere, the $U(1)$ gauge symmetry and the $U(1)_R$ symmetry.}

As in flat space, the transformation rules (\ref{var-s^3}) are not purely SUSY variations, since they are composed of SUSY variation plus gauge transformation with an appropriately chosen gauge parameter, such that the modified variations do not take us out of the gauge choice. Thus, these variations do not close on the usual SUSY algebra:
\begin{equation}
\begin{aligned}
&{[\delta_\xi,\delta_{\bar\xi}]}\phi\;\;=-2i(\xi\sigma^\mu\bar\xi)D_\mu\phi\;,\\
&[\delta_\xi,\delta_{\bar\xi}]\psi_\alpha=-2i[(\xi\sigma^\mu\bar\xi)(D_\mu{\psi})_\alpha-{1\over 2}\nabla_\mu(\xi\sigma_\nu\bar\xi)\sigma^{\mu\nu}\psi]\;,\\
&[\delta_\xi,\delta_{\bar\xi}]F\;\;=-2i(\xi\sigma^\mu\bar\xi)D_\mu{F}\;,\\
&[\delta_\xi,\delta_{\bar\xi}]A_{\mu}=-2i(\xi\sigma^\nu\bar\xi)F_{\nu\mu}\;,\\
&[\delta_\xi,\delta_{\bar\xi}]\lambda_\alpha=-2i[(\xi\sigma^\mu\bar\xi)(D_\mu{\lambda})_\alpha-{1\over 2}\nabla_\mu(\xi\sigma_\nu\bar\xi)\sigma^{\mu\nu}\lambda]\;,\\
&[\delta_\xi,\delta_{\bar\xi}]D\;=-2i(\xi\sigma^\mu\bar\xi)D_\mu{D}\;.
\end{aligned}\label{modified-algebra}
\end{equation}
Recalling that the gauge transformations of the fields are:
\begin{align*}
&\delta^G_\alpha\phi\;=ig\alpha\phi\;\;\;,\;\;\;\delta^G_\alpha\psi=ig\alpha\psi\;\;\;,\;\;\;\delta^G_\alpha{F}=ig\alpha{F}\;,\\
&\delta^G_\alpha{A_\mu}=\partial_\mu\alpha\;\;\;,\;\;\;\delta^G_\alpha\lambda=0\;\;\;\;\;\;\;\;\;,\;\;\;\delta^G_\alpha{D}=0\;,
\end{align*}
and that the Lie-derivatives of the fields are:
\begin{equation*}
\begin{aligned}
&{\cal{L}}_v\{\phi,F,D\}=v_\mu\partial^\mu\{\phi,F,D\}\;,\\
&{\cal{L}}_v\{\psi,\lambda\}=v_\mu\nabla^\mu\{\psi,\lambda\}-{1\over2}(\nabla_\mu{v_\nu})\sigma^{\mu\nu}\{\psi,\lambda\}\;,\\
&{\cal{L}}_v{A_\mu}=v^\nu\nabla_\nu{A_\mu}+(\nabla_\mu{v^\nu}){A_\nu}\;,
\end{aligned}
\end{equation*}
we find that
\begin{equation*}[\delta_\xi,\delta_{\bar\xi}]=-2i({\cal{L}}_{v}-{v^\mu{\tilde{A_\mu}}}\delta^R-\delta^G_{v^\mu{A_\mu}})\;,\end{equation*}
where:
\begin{equation*}v^\mu=\xi\sigma^\mu\bar\xi\;,\end{equation*}
and $\delta^Ry$ is the R-variation of the field y (e.g. $\delta^R\phi=iq\phi$).

\subsection{Mode Expansion}
We consider the gauge sector of the theory defined above, and eliminate the residual gauge invariance by fixing $A_0=0$. This gauge choice does not require any further modification of the SUSY transformation laws (besides from setting $A_0=0$ everywhere).

We expand the fields in spherical harmonics:
\begin{align}\lambda&=\sum{\lambda^\epsilon_{k,m,n}(t)\psi^\epsilon_{k,m,n}(\theta,\alpha,\beta)}\;,\label{expansion1}\\
\vec{A}&=\sum{A^\epsilon_{k,m,n}(t)\vec{V}^\epsilon_{k,m,n}}(\theta,\alpha,\beta)\;.\label{expansion2}\end{align}
Here $\vec{V}^\epsilon_{k,m,n}$ are the vector spherical harmonics. Some of the useful properties of these functions are summarized in Appendix \ref{spherical_harmonics} (in general, the expansion of a vector field should also include components proportional to $\vec\nabla{\phi_{k,m,n}}$ where $\phi_{k,m,n}$ are the scalar spherical harmonics, but these components are eliminated by our gauge choice).

The Lagrangian can be written in terms of these modes as:
\begin{align*}
{{L}}_{gauge}=&\sum_{k,m,n,\epsilon}{1\over2}(-1)^{{{m+n}\over2}+1}[(\partial_0A^{\epsilon}_{k,m,n})(\partial_0A^{\epsilon}_{k,-m,-n})-{{(k+1+\epsilon)^2}\over{r^2}}A^\epsilon_{k,m,n}A^\epsilon_{k,-m,-n}]
\\+&\sum_{k,m,n,\epsilon}i\lambda^{*\epsilon}_{k,m,n}\left(\partial_0+{i\over{2r}}+{{i\epsilon}\over{r}}(k+{{2+\epsilon}\over{2}})\right)\lambda^\epsilon_{k,m,n}\;.
\end{align*}

By setting $\xi^1=0$ in the Noether procedure we pick the supercharge $Q_2$.\footnote{In order to calculate $Q_2$ in terms of the modes we had to calculate an integral of the product of 2 spinor and one vector spherical harmonics (from the mode expansion of the gaugino, the SUSY spinor and the gauge vector), we used the formula (\ref{formula}) for integrals of this type, which was obtained in \cite{Mussel:2009uw}. We have found:
\begin{equation}
\begin{aligned}
Q_2&=\sqrt2\pi{r^{3\over2}}\int_{S^3}{d{\theta}d{\alpha}d{\beta}\sqrt{g}[\sum_{k,m,n,\epsilon}\sum_{k',m',n',{\epsilon'}}\left(i\partial_0-{\epsilon \over r}(k+1+\epsilon)\right)A^\epsilon_{k,m,n}\lambda^{*{\epsilon'}}_{k',m',n'}V^{\epsilon,i}_{k,m,n}(\psi^{\epsilon'}_{k',m',n'})^\dagger\bar\sigma_i\psi^{1}_{0,-1,0}]}\\&=\sum_{k,m,n,\epsilon}{{(-1)^{{m(1-\epsilon)}\over{2}}}}\sqrt{{k+\epsilon{m}+1+\epsilon}\over{k+1+\epsilon}}\left((-1)^{{{m+n}\over2}+1}\Pi_{A^\epsilon_{k,-m,-n}}+i{\epsilon \over r}(k+1+\epsilon)A^\epsilon_{k,m,n}\right)\lambda^{*{\epsilon}}_{k,m-1,n}\;.
\end{aligned}\label{integral}
\end{equation}}
The operator $\Delta_2=~{1\over2}\{Q_2,Q^\dagger_2\}$ is given by:\footnote{The sums here and in (\ref{integral},\ref{delta_2_gauge}) are over the indices corresponding to the expansion in vector spherical harmonics, i.e:
\begin{equation*}
\begin{aligned}
k\geq0\;\;\;,\;\;\;|m|\leq{k+2}\;\;\;,\;\;\;|n|\leq{k}\;\;\;,\;\;\;m \equiv n \equiv k \pmod 2\;\;\;&\text{for } \epsilon=+1\;,\\
k\geq2\;\;\;,\;\;\;|m|\leq{k-2}\;\;\;,\;\;\;|n|\leq{k}\;\;\;,\;\;\;m \equiv n \equiv k \pmod 2\;\;\;&\text{for } \epsilon=-1\;.
\end{aligned}
\end{equation*}
The conjugate modes are:
\begin{equation*}
\Pi_{A^\epsilon_{k,m,n}}=(-1)^{{{m+n}\over2}+1}\partial_0A^\epsilon_{k,-m,-n}\;\;\;,\;\;\;
\Pi_{\lambda^\epsilon_{k,m,n}}=-i\lambda^{\epsilon*}_{k,m,n}\;.
\end{equation*}}\footnote{The bosonic part of $\Delta_2$ can be written as a sum of ``Landau-like" Hamiltonians, using $(A^\epsilon_{k,m,n})^*=(-1)^{{{m+n}\over2}+1}A^\epsilon_{k,-m,-n}$.}
\begin{align*}
\Delta_2=&\sum_{\epsilon,k,m,n}[{1\over2}(-1)^{{{m+n}\over2}+1}\Pi_{A^\epsilon_{k,m,n}}\Pi_{A^\epsilon_{k,-m,-n}}+{1\over2}{{(k+1+\epsilon)^2}\over{r^2}}(-1)^{{{m+n}\over2}+1}{A^\epsilon_{k,m,n}}{A^\epsilon_{k,-m,-n}}\\
&+i{{k+\epsilon{m}+1+\epsilon}\over{2}}{\epsilon\over{r}}\left(\Pi_{A^\epsilon_{k,m,n}}A^\epsilon_{k,m,n}-A^\epsilon_{k,-m,-n}\Pi_{A^\epsilon_{k,-m,-n}}\right)-{\epsilon\over{r}}{{(k+\epsilon{m}+1+\epsilon)}}\lambda^{*\epsilon}_{k,m-1,n}\lambda^{\epsilon}_{k,m-1,n}]\;.
\end{align*}

\subsection{Reproducing the Index}
In order to diagonalize $\Delta_2$ we define creation and annihilation operators,
\begin{align*}
C^\epsilon_{k,m,n}&=\sqrt{r\over{2(1+k+\epsilon)}}\left(\Pi_{A^\epsilon_{k,m,n}}-i{{(1+k+\epsilon)}\over{r}}(-1)^{{{m+n}\over2}+1}A^\epsilon_{k,-m,-n}\right)\;,\\
C^{\epsilon\dagger}_{k,m,n}&=\sqrt{r\over{2(1+k+\epsilon)}}\left((-1)^{{{m+n}\over2}+1}\Pi_{A^\epsilon_{k,-m,-n}}+i{{(1+k+\epsilon)}\over{r}}A^\epsilon_{k,m,n}\right)\;,\\
[C^\epsilon_{k,m,n}&,C^{\epsilon\dagger}_{k,m,n}]=1\;,
\end{align*}
in terms of which we get:
\begin{align}
\Delta_2=\sum_{k,m,n,\epsilon}{{k+1+\epsilon+m}\over{r}}C^{\epsilon\dagger}_{k,m,n}C^{\epsilon}_{k,m,n}+\sum_{k,m,n}{{k+2+m}\over{r}}\lambda^{1}_{k,m-1,n}\lambda^{1\dagger}_{k,m-1,n}+\sum_{k,m,n}{{k-m}\over{r}}\lambda^{-1\dagger}_{k,m-1,n}\lambda^{-1}_{k,m-1,n}\;.\label{delta_2_gauge}
\end{align}
The contributions to the index come from the kernel of $\Delta_2$, which is generated by the creation operators $C^{1\dagger}_{k,-(k+2),n}$ and $\lambda^{-1\dagger}_{k,k-1,n}$.
Thus the bosonic ``single particle" zero modes have eigenvalues:\footnote{\begin{align*}
J^3_L&=-\sum_{k,m,n,\epsilon}mC^{\epsilon\dagger}_{k,m,n}C^{\epsilon}_{k,m,n}-\sum_{k,m,n}m\lambda^1_{k,m,n}\lambda^{1*}_{k,m,n}+\sum_{k,m,n}m\lambda^{-1*}_{k,m,n}\lambda^{-1}_{k,m,n}\;,\\
J^3_R&=-\sum_{k,m,n,\epsilon}nC^{\epsilon\dagger}_{k,m,n}C^{\epsilon}_{k,m,n}-\sum_{k,m,n}n\lambda^1_{k,m,n}\lambda^{1*}_{k,m,n}+\sum_{k,m,n}n\lambda^{-1*}_{k,m,n}\lambda^{-1}_{k,m,n}\;,\\
R&=-\sum_{k,m,n}\lambda^1_{k,m,n}\lambda^{1*}_{k,m,n}+\sum_{k,m,n}\lambda^{-1*}_{k,m,n}\lambda^{-1}_{k,m,n}\;.
\end{align*}}
\begin{align*}
R+J^3_L&=k+2\;,\\J^3_R&=-n\;,
\end{align*}
and for the fermionic zero modes:
\begin{align*}
R+J^3_L&=k\;,\\J^3_R&=n\;.
\end{align*}
The generating function is therefore:
\begin{align*}
&\sum_{k=0}^{\infty}t^{k+2}\sum_ny^{-n}-\sum_{k=1}^{\infty}t^k\sum_ny^n={{2t^2-t(y+{1\over{y}})}\over{(1-{t\over{y}})(1-ty)}}\;,
\end{align*}
in agreement with \cite{Romelsberger:2007ec}.

\newpage

\bibliography{thesisbib}

\providecommand{\href}[2]{#2}\begingroup\raggedright\begin{thebibliography}{10}

\bibitem{Romelsberger:2005eg}
C.~Romelsberger, ``{Counting chiral primaries in N = 1, d=4 superconformal
  field theories},''
  \href{http://dx.doi.org/10.1016/j.nuclphysb.2006.03.037}{{\em Nucl.Phys.}
  {\bfseries B747} (2006) 329--353},
\href{http://arxiv.org/abs/hep-th/0510060}{{\ttfamily arXiv:hep-th/0510060
  [hep-th]}}.

\bibitem{Kinney:2005ej}
J.~Kinney, J.~M. Maldacena, S.~Minwalla, and S.~Raju, ``{An Index for 4
  dimensional super conformal theories},''
  \href{http://dx.doi.org/10.1007/s00220-007-0258-7}{{\em Commun.Math.Phys.}
  {\bfseries 275} (2007) 209--254},
\href{http://arxiv.org/abs/hep-th/0510251}{{\ttfamily arXiv:hep-th/0510251
  [hep-th]}}.

\bibitem{Pestun:2007rz}
V.~Pestun, ``{Localization of gauge theory on a four-sphere and supersymmetric
  Wilson loops},'' \href{http://dx.doi.org/10.1007/s00220-012-1485-0}{{\em
  Commun.Math.Phys.} {\bfseries 313} (2012) 71--129},
\href{http://arxiv.org/abs/0712.2824}{{\ttfamily arXiv:0712.2824 [hep-th]}}.

\bibitem{Dolan:2008qi}
F.~Dolan and H.~Osborn, ``{Applications of the Superconformal Index for
  Protected Operators and q-Hypergeometric Identities to N=1 Dual Theories},''
  \href{http://dx.doi.org/10.1016/j.nuclphysb.2009.01.028}{{\em Nucl.Phys.}
  {\bfseries B818} (2009) 137--178},
\href{http://arxiv.org/abs/0801.4947}{{\ttfamily arXiv:0801.4947 [hep-th]}}.

\bibitem{Closset:2012vg}
C.~Closset, T.~T. Dumitrescu, G.~Festuccia, Z.~Komargodski, and N.~Seiberg,
  ``{Contact Terms, Unitarity, and F-Maximization in Three-Dimensional
  Superconformal Theories},''
  \href{http://dx.doi.org/10.1007/JHEP10(2012)053}{{\em JHEP} {\bfseries 1210}
  (2012) 053},
\href{http://arxiv.org/abs/1205.4142}{{\ttfamily arXiv:1205.4142 [hep-th]}}.

\bibitem{Closset:2013vra}
C.~Closset, T.~T. Dumitrescu, G.~Festuccia, and Z.~Komargodski, ``{The Geometry
  of Supersymmetric Partition Functions},''
\href{http://arxiv.org/abs/1309.5876}{{\ttfamily arXiv:1309.5876 [hep-th]}}.

\bibitem{Festuccia:2011ws}
G.~Festuccia and N.~Seiberg, ``{Rigid Supersymmetric Theories in Curved
  Superspace},'' \href{http://dx.doi.org/10.1007/JHEP06(2011)114}{{\em JHEP}
  {\bfseries 1106} (2011) 114},
\href{http://arxiv.org/abs/1105.0689}{{\ttfamily arXiv:1105.0689 [hep-th]}}.

\bibitem{Dumitrescu:2012ha}
T.~T. Dumitrescu, G.~Festuccia, and N.~Seiberg, ``{Exploring Curved
  Superspace},'' \href{http://dx.doi.org/10.1007/JHEP08(2012)141}{{\em JHEP}
  {\bfseries 1208} (2012) 141},
\href{http://arxiv.org/abs/1205.1115}{{\ttfamily arXiv:1205.1115 [hep-th]}}.

\bibitem{Dumitrescu:2012at}
T.~T. Dumitrescu and G.~Festuccia, ``{Exploring Curved Superspace (II)},''
  \href{http://dx.doi.org/10.1007/JHEP01(2013)072}{{\em JHEP} {\bfseries 1301}
  (2013) 072},
\href{http://arxiv.org/abs/1209.5408}{{\ttfamily arXiv:1209.5408 [hep-th]}}.

\bibitem{Romelsberger:2007ec}
C.~Romelsberger, ``{Calculating the Superconformal Index and Seiberg
  Duality},''
\href{http://arxiv.org/abs/0707.3702}{{\ttfamily arXiv:0707.3702 [hep-th]}}.

\bibitem{Spiridonov:2008zr}
V.~Spiridonov and G.~Vartanov, ``{Superconformal indices for N = 1 theories
  with multiple duals},''
  \href{http://dx.doi.org/10.1016/j.nuclphysb.2009.08.022}{{\em Nucl.Phys.}
  {\bfseries B824} (2010) 192--216},
\href{http://arxiv.org/abs/0811.1909}{{\ttfamily arXiv:0811.1909 [hep-th]}}.

\bibitem{Spiridonov:2009za}
V.~Spiridonov and G.~Vartanov, ``{Elliptic Hypergeometry of Supersymmetric
  Dualities},'' \href{http://dx.doi.org/10.1007/s00220-011-1218-9}{{\em
  Commun.Math.Phys.} {\bfseries 304} (2011) 797--874},
\href{http://arxiv.org/abs/0910.5944}{{\ttfamily arXiv:0910.5944 [hep-th]}}.

\bibitem{Spiridonov:2011hf}
V.~Spiridonov and G.~Vartanov, ``{Elliptic hypergeometry of supersymmetric
  dualities II. Orthogonal groups, knots, and vortices},''
\href{http://arxiv.org/abs/1107.5788}{{\ttfamily arXiv:1107.5788 [hep-th]}}.

\bibitem{Gadde:2010en}
A.~Gadde, L.~Rastelli, S.~S. Razamat, and W.~Yan, ``{On the Superconformal
  Index of N=1 IR Fixed Points: A Holographic Check},''
  \href{http://dx.doi.org/10.1007/JHEP03(2011)041}{{\em JHEP} {\bfseries 1103}
  (2011) 041},
\href{http://arxiv.org/abs/1011.5278}{{\ttfamily arXiv:1011.5278 [hep-th]}}.

\bibitem{spiridonov2001elliptic}
V.~P. Spiridonov, ``{On the elliptic beta function},'' {\em Russian
  Mathematical Surveys} {\bfseries 56} no.~1, (2001) 185--186.

\bibitem{spiridonov2004theta}
V.~Spiridonov, ``Theta hypergeometric integrals,'' {\em St. Petersburg
  Mathematical Journal} {\bfseries 15} no.~6, (2004) 929--967.

\bibitem{spiridonov2008essays}
V.~P. Spiridonov, ``Essays on the theory of elliptic hypergeometric
  functions,'' {\em Russian Mathematical Surveys} {\bfseries 63} no.~3, (2008)
  405.

\bibitem{rains2003transformations}
E.~M. Rains, ``{Transformations of elliptic hypergometric integrals},'' {\em
  arXiv preprint math/0309252} (2003) .

\bibitem{Spiridonov:2010hh}
V.~Spiridonov and G.~Vartanov, ``{Supersymmetric dualities beyond the conformal
  window},'' \href{http://dx.doi.org/10.1103/PhysRevLett.105.061603}{{\em
  Phys.Rev.Lett.} {\bfseries 105} (2010) 061603},
\href{http://arxiv.org/abs/1003.6109}{{\ttfamily arXiv:1003.6109 [hep-th]}}.

\bibitem{Gadde:2009kb}
A.~Gadde, E.~Pomoni, L.~Rastelli, and S.~S. Razamat, ``{S-duality and 2d
  Topological QFT},'' \href{http://dx.doi.org/10.1007/JHEP03(2010)032}{{\em
  JHEP} {\bfseries 1003} (2010) 032},
\href{http://arxiv.org/abs/0910.2225}{{\ttfamily arXiv:0910.2225 [hep-th]}}.

\bibitem{Spiridonov:2010qv}
V.~Spiridonov and G.~Vartanov, ``{Superconformal indices of ${\mathcal N}=4$
  SYM field theories},''
  \href{http://dx.doi.org/10.1007/s11005-011-0537-2}{{\em Lett.Math.Phys.}
  {\bfseries 100} (2012) 97--118},
\href{http://arxiv.org/abs/1005.4196}{{\ttfamily arXiv:1005.4196 [hep-th]}}.

\bibitem{Witten:1982df}
E.~Witten, ``{Constraints on Supersymmetry Breaking},''
\href{http://dx.doi.org/10.1016/0550-3213(82)90071-2}{{\em Nucl.Phys.}
  {\bfseries B202} (1982) 253}.

\bibitem{Intriligator:2005if}
K.~A. Intriligator, ``{IR free or interacting? A Proposed diagnostic},''
  \href{http://dx.doi.org/10.1016/j.nuclphysb.2005.10.005}{{\em Nucl.Phys.}
  {\bfseries B730} (2005) 239--251},
\href{http://arxiv.org/abs/hep-th/0509085}{{\ttfamily arXiv:hep-th/0509085
  [hep-th]}}.

\bibitem{Intriligator:2003jj}
K.~A. Intriligator and B.~Wecht, ``{The Exact superconformal R symmetry
  maximizes a},'' \href{http://dx.doi.org/10.1016/S0550-3213(03)00459-0}{{\em
  Nucl.Phys.} {\bfseries B667} (2003) 183--200},
\href{http://arxiv.org/abs/hep-th/0304128}{{\ttfamily arXiv:hep-th/0304128
  [hep-th]}}.

\bibitem{Intriligator:1994rx}
K.~A. Intriligator, N.~Seiberg, and S.~Shenker, ``{Proposal for a simple model
  of dynamical SUSY breaking},''
  \href{http://dx.doi.org/10.1016/0370-2693(94)01336-B}{{\em Phys.Lett.}
  {\bfseries B342} (1995) 152--154},
\href{http://arxiv.org/abs/hep-ph/9410203}{{\ttfamily arXiv:hep-ph/9410203
  [hep-ph]}}.

\bibitem{Brodie:1998vv}
J.~H. Brodie, P.~L. Cho, and K.~A. Intriligator, ``{Misleading anomaly
  matchings?},'' \href{http://dx.doi.org/10.1016/S0370-2693(98)00353-0}{{\em
  Phys.Lett.} {\bfseries B429} (1998) 319--326},
\href{http://arxiv.org/abs/hep-th/9802092}{{\ttfamily arXiv:hep-th/9802092
  [hep-th]}}.

\bibitem{Poppitz:2009kz}
E.~Poppitz and M.~Unsal, ``{Chiral gauge dynamics and dynamical supersymmetry
  breaking},'' \href{http://dx.doi.org/10.1088/1126-6708/2009/07/060}{{\em
  JHEP} {\bfseries 0907} (2009) 060},
\href{http://arxiv.org/abs/0905.0634}{{\ttfamily arXiv:0905.0634 [hep-th]}}.

\bibitem{Vartanov:2010xj}
G.~Vartanov, ``{On the ISS model of dynamical SUSY breaking},''
  \href{http://dx.doi.org/10.1016/j.physletb.2010.12.040}{{\em Phys.Lett.}
  {\bfseries B696} (2011) 288--290},
\href{http://arxiv.org/abs/1009.2153}{{\ttfamily arXiv:1009.2153 [hep-th]}}.

\bibitem{Buican:2011ty}
M.~Buican, ``{A Conjectured Bound on Accidental Symmetries},''
  \href{http://dx.doi.org/10.1103/PhysRevD.85.025020}{{\em Phys.Rev.}
  {\bfseries D85} (2012) 025020},
\href{http://arxiv.org/abs/1109.3279}{{\ttfamily arXiv:1109.3279 [hep-th]}}.

\bibitem{Benvenuti:2006qr}
S.~Benvenuti, B.~Feng, A.~Hanany, and Y.-H. He, ``{Counting BPS Operators in
  Gauge Theories: Quivers, Syzygies and Plethystics},''
  \href{http://dx.doi.org/10.1088/1126-6708/2007/11/050}{{\em JHEP} {\bfseries
  0711} (2007) 050},
\href{http://arxiv.org/abs/hep-th/0608050}{{\ttfamily arXiv:hep-th/0608050
  [hep-th]}}.

\bibitem{Feng:2007ur}
B.~Feng, A.~Hanany, and Y.-H. He, ``{Counting gauge invariants: The Plethystic
  program},'' \href{http://dx.doi.org/10.1088/1126-6708/2007/03/090}{{\em JHEP}
  {\bfseries 0703} (2007) 090},
\href{http://arxiv.org/abs/hep-th/0701063}{{\ttfamily arXiv:hep-th/0701063
  [hep-th]}}.

\bibitem{Wess:1992cp}
J.~Wess and J.~Bagger,
``{Supersymmetry and supergravity},''.

\bibitem{Mussel:2009uw}
M.~Mussel and R.~Yacoby, ``{The 2-loop partition function of large N gauge
  theories with adjoint matter on S**3},''
  \href{http://dx.doi.org/10.1088/1126-6708/2009/12/005}{{\em JHEP} {\bfseries
  0912} (2009) 005},
\href{http://arxiv.org/abs/0909.0407}{{\ttfamily arXiv:0909.0407 [hep-th]}}.

\bibitem{Aharony:2005bq}
O.~Aharony, J.~Marsano, S.~Minwalla, K.~Papadodimas, and M.~Van~Raamsdonk, ``{A
  First order deconfinement transition in large N Yang-Mills theory on a small
  S**3},'' \href{http://dx.doi.org/10.1103/PhysRevD.71.125018}{{\em Phys.Rev.}
  {\bfseries D71} (2005) 125018},
\href{http://arxiv.org/abs/hep-th/0502149}{{\ttfamily arXiv:hep-th/0502149
  [hep-th]}}.

\bibitem{Sohnius:1982fw}
M.~Sohnius and P.~C. West, ``{The tensor calculus and matter coupling of the
  alternative minimal auxiliary field formulation of N=1 supergravity},''
\href{http://dx.doi.org/10.1016/0550-3213(82)90337-6}{{\em Nucl.Phys.}
  {\bfseries B198} (1982) 493}.

\end{thebibliography}\endgroup
\bibliographystyle{utphys}

\end{document}